\documentclass{aa}  
\usepackage{graphicx}
\usepackage{txfonts}
\usepackage{lipsum}
\usepackage{subcaption} 
\usepackage{lscape} 
\usepackage{placeins}         
\usepackage{xcolor}
\definecolor{cobalt}{rgb}{0.2118, 0.3176, 0.5804}
\usepackage{hyperref}    
\hypersetup{
  colorlinks=true,
  citecolor=cobalt,
  linkcolor=cobalt,
  urlcolor=cobalt,
}
\usepackage[bottom]{footmisc}
\usepackage{float}
\usepackage{stfloats}
\usepackage{cleveref} 
\Crefname{equation}{Eq.}{Eqs.}
\Crefname{figure}{Fig.}{Figs.}
\usepackage{orcidlink}
\usepackage{siunitx}
\usepackage{multirow}
\usepackage{comment}
\usepackage{caption}
\def\dd{\mathrm{d}}
\def\l{\left}
\def\r{\right}
\def\f{\frac}
\def\chimera{\texttt{CHIMERA}\xspace}

\def\pltwop{\textsc{Power Law + Double Peak}\xspace}
\def\bspline{\textsc{BSpline}\xspace}
\def\bsplines{\textsc{BSplines}\xspace}
\DeclareSIUnit \parsec {pc}
\DeclareSIUnit \years{yrs}
\newcommand{\msun}{\ensuremath{\mathrm{M}_\odot}}

\begin{document}

\title{Mind the peak: improving cosmological constraints from GWTC-4.0 spectral sirens using semiparametric mass models}

\author{Matteo Tagliazucchi\inst{1,2,3}\corrauth{\href{emailto:matteo.tagliazucchi2@unibo.it}{matteo.tagliazucchi2@unibo.it}}\orcidlink{0009-0003-8886-3184} 
\and Michele Moresco\inst{1,2}\orcidlink{0000-0002-7616-7136} 
\and Nicola Borghi\inst{1,2,3}\orcidlink{0000-0002-2889-8997} 
\and Chiara Ciapetti\inst{1}
}

\institute{Dipartimento di Fisica e Astronomia ``Augusto Righi''--Universit\`{a} di Bologna, via Piero Gobetti 93/2, I-40129 Bologna, Italy
\and
INAF - Osservatorio di Astrofisica e Scienza dello Spazio di Bologna, via Piero Gobetti 93/3, I-40129 Bologna, Italy
\and
INFN - Sezione di Bologna, Viale Berti Pichat 6/2, I-40127 Bologna, Italy
}

\date{Received: 23 December 2025 / Accepted: 20 April 2026} 

\abstract{
    Gravitational wave spectral sirens can provide cosmological constraints by using the shape of the binary black hole (BBH) mass distribution (MD). However, the precision and accuracy of these constraints depend critically on the capturing all the MD features. In this work, we analyzed 137 BBH events from the latest GWTC-4.0 with a novel data-driven semiparametric approach based on \textsc{Bspline} that adaptively places knots around the most informative structures in the MD, while keeping the dimensionality of the parameter space moderate. Our flexible models resolved three distinct peaks at $\sim10$, $18$, and $33\,\mathrm{M}_\odot$ and are statistically preferred over standard parametric models, with Bayes factors up to 226. 
    Because these features are correlated with $H_0$, the semiparametric model yielded, under different prior assumptions, 12\%-21\% improvement in the precision of $H_0$ relative to parametric models, providing $H_0 = 57.8^{+21.9}_{-20.6}\,\mathrm{km/s/Mpc}$ in the best case. 
    Our results demonstrate that capturing the full complexity of the BBH mass distribution is essential for realizing the cosmological potential of spectral sirens as gravitational wave catalogs continue to grow.
}


\titlerunning{Improving GWTC-4.0 spectral siren cosmological constraints using semiparametric mass models}

\maketitle
\nolinenumbers

\section{Introduction}
Since their first discovery 10 years ago \citep{LIGOScientific:2016aoc}, gravitational-wave (GW) events from compact binary systems have proven to be an extremely promising cosmological probe for directly measuring the expansion rate of the Universe \citep[standard sirens,][]{Schutz:1986gp}.
By combining the luminosity distance from GWs with redshift information, they offer an independent test to address the tension between local and early-Universe measurements of the Hubble constant \citep[$H_0$; e.g., see][]{Moresco:2022phi, Jin:2025dvf}.
In this work, we focus on the ``spectral siren'' method, which extracts redshift information by statistically breaking the mass-redshift degeneracy using features in the source-frame mass distribution (MD) of compact binaries \citep{Chernoff:1993th, Taylor:2011fs, Mancarella:2021ecn, Ezquiaga:2022zkx, Chen:2024gdn, Mali:2024wpq}. 
A critical aspect of this technique is the accurate modeling of the MD, as incorrect assumptions and simplified templates can introduce biases in the inferred $H_0$ and reduce the constraining power of this method \citep[e.g.,][]{Pierra:2023deu, Agarwal:2024hld}.
So far, most of the mass modeling for spectral siren cosmology has followed a template-driven approach in which a parametric form is assumed and eventually revised as more observations are made. 
For instance, the baseline binary black hole (BBH) MD adopted in the GWTC-3 cosmological analysis was a power law plus a single Gaussian peak \citep{LIGOScientific:2021aug}, whereas recent results from GWTC-4 are obtained using a model with two Gaussian peaks \citep{LIGOScientific:2025jau}. 
While flexible, these parametric models may not capture the full complexity of BBH MD.
Moreover, guessing the correct MD shape in advance can be a challenge for the spectral siren method.
We address these limitations by using a semiparametric model, similar to the one used in population studies by \cite{Edelman:2021zkw}, that is sufficiently flexible to capture unknown substructures in the BBH MD.
Recent nonparametric studies have employed Gaussian processes for the primary mass distribution \citep{Farah:2024xub, MaganaHernandez:2025cnu} and \bspline for cosmic expansion $H(z)$ \citep{Pierra:2025hoc}.
Employing \bspline basis functions together with a novel data-driven optimization of knot positions, we demonstrate that a more detailed reconstruction of the BBH MD enables a significantly more powerful extraction of cosmological information.
We include this model within the \chimera pipeline \citep{Borghi:2023opd, Tagliazucchi:2025ofb, Borghi:2025pav}, which we use to jointly infer $H_0$ and population parameters from a subset of GWTC-4.0 BBHs.

\section{Data}

We analyzed the same dataset adopted in the GWTC-4.0 cosmology paper \citep{LIGOScientific:2025jau}, consisting of 137 BBHs detected between the first and the first part of the fourth observing run (O1 to O4a) of the LIGO-Virgo-KAGRA (LVK) collaboration \citep{KAGRA:2013rdx, LIGOScientific:2014pky, VIRGO:2014yos, KAGRA:2020tym}, with a false alarm rate (FAR) of less than 0.25 per year.
In particular, three of these BBHs are from the O1, seven from O2, 52 from O3, and 75 from O4a \citep{LIGOScientific:2018mvr, LIGOScientific:2021usb, KAGRA:2021vkt, LIGOScientific:2025slb}. 
This dataset conservatively excludes GW231123\_135430, as its extreme properties (high spins and mass) push waveform models beyond their well-calibrated regime \citep{LIGOScientific:2025rsn}. 
To align with the assumption of \citet{LIGOScientific:2025jau} about the dataset used, we also exclude GW190814 as a possible neutron star-black hole merger.
We approximate the source property distributions for all events using 5000 posterior-estimate (PE) samples obtained with a single waveform model, consistent with the approach of \citep{LIGOScientific:2025jau} to minimize waveform systematics. For the O1–O3 events, we use thed \textsc{IMRPhenomXPHM} model, while for O4a events we use its updated version, \textsc{IMRPhenomXPHM\_SpinTaylor}.
We used the PE samples publicly released by the LVK collaboration \citep{LIGOScientific:2022gwtc2, LIGOScientific:2023gwtc3, LIGOScientific:2025gwtc4}. 
The injections used to account for selection effects are from the public O3–O4a set in \cite{LIGOScientific:2025yae, LIGOScientific:2025gwtc4-injections}, with O1–O2 search sensitivity estimate handled via a semi-analytic model \citep{Essick:2023toz}. 
We marginalized over spin parameters, as they are not included in this analysis.

\section{Methods}

We use a hierarchical Bayesian framework \citep{Mandel:2018mve, Vitale:2020aaz, Gair:2022zsa} to infer cosmological and population hyperparameters from GW data.
The framework employs a hyper-likelihood that describes the probability of observing the data $\{\boldsymbol{d}_i\}$ given an astrophysical population model $p_{\rm pop}(\boldsymbol{\theta} \mid \boldsymbol{\Lambda})$:
\begin{align}
    \mathcal{L}\l(\{\boldsymbol{d}_i\} \mid \boldsymbol{\Lambda}\r) \propto \prod_{i=1}^{N_{\rm obs}}\f{ \int \dd \boldsymbol{\theta}_{\dd,i}  \mathcal{L}_{\rm gw}\l( \boldsymbol{d}_i  \mid  \boldsymbol{\theta}_{\dd,i}\r)  \left|\f{\dd \boldsymbol{\theta}_i}{\dd \boldsymbol{\theta}_{\dd,i}} \right| p_{\rm pop}(\boldsymbol{\theta}_i \mid \boldsymbol{\Lambda}) }{\int \dd \boldsymbol{\theta}_\dd P_{\rm det} (\boldsymbol{\theta}_\dd) \l|\f{\dd \boldsymbol{\theta}}{\dd \boldsymbol{\theta}_\dd} \r|  p_{\rm pop}(\boldsymbol{\theta} \mid \boldsymbol{\Lambda})   } \label{eq:hierarchical-lkl} \;.
\end{align}
Here, $\boldsymbol{\theta}_\dd$ represents the GW source parameters in detector-frame, such as luminosity distance and redshifted binary masses, while $\boldsymbol{\theta}$ represents the corresponding parameters in the source frame.
The Jacobian term,  $\left|\f{\dd \boldsymbol{\theta}_i}{\dd \boldsymbol{\theta}_{\dd,i}} \right|$, converts the population prior, $p_{\rm pop}$, from the source frame to the detector frame.
The population prior describes the probability of drawing a GW event with source parameters $\boldsymbol{\theta}_i$ from a population described by hyperparameters $\boldsymbol{\Lambda}$.
The $P_{\rm det}(\boldsymbol{\theta}_\dd)$ term appearing in \Cref{eq:hierarchical-lkl} is the probability of detecting a GW source with parameters $\boldsymbol{\theta}_\dd$.
Overall, the denominator of \Cref{eq:hierarchical-lkl} is the fraction of GW sources drawn from the population modeled by $p_{\rm pop}$ that can be detected by the considered network of interferometers, and is evaluated using an injection campaign as described in \cite{Borghi:2023opd,Tagliazucchi:2025ofb}.

In this work, we considered the ``spectral sirens'' method and we neglected spins. 
In this case the population prior is factorized as
\begin{equation}
    p_{\rm pop}(\boldsymbol{\theta}\mid \boldsymbol{\Lambda}) \propto p(m_1, m_2 \mid \boldsymbol{\lambda}_m) \psi(z \mid \boldsymbol{\lambda}_r) \f{\dd V_c}{\dd z}(z\mid \boldsymbol{\lambda}_c),
\end{equation}
where $\f{\dd V_c}{\dd z}(z\mid \boldsymbol{\lambda}_c)$ is the comoving volume element parametrized by cosmological parameters $\boldsymbol{\lambda}_c$, $\psi(z\mid\boldsymbol{\lambda}_r)$ describes the redshift evolution of the merger rate, and $p(m_1, m_2 \boldsymbol{\lambda}_m)$ is the GW mass distribution.
We modeled the merger rate evolution using the Madau-Dickinson law as in \citet{LIGOScientific:2025jau}:
\begin{equation}
    \psi(z \mid \boldsymbol{\lambda}_r) \propto \f{(1+z)^\gamma}{1+\l(\f{1+z}{1+z_p}\r)^{\gamma+\kappa}}.
\end{equation}

The mass distribution, which includes the spectral features that are exploited to break the mass-redshift degeneracy and infer cosmological parameters, is factorized as 
\begin{equation}
    p(m_1, m_2 \mid \boldsymbol{\lambda}_m ) = p(m_1 \mid \boldsymbol{\lambda}_m ) p(m_2 \mid m_1, \boldsymbol{\lambda}_m ).
\end{equation}
In this work, we considered two different functional forms of the primary mass distribution, $p(m_1 \mid \boldsymbol{\lambda_m})$, a parametric and a semiparametric one.
Both models assume a non-evolving mass function in redshift.
This assumption, justified by the limited redshift horizon of current detectors, is made to align with that of \cite{LIGOScientific:2025jau} and thus ensure a fair comparison of the cosmological constraints.
The parametric function is the \pltwop (\textsc{pl2p}), consisting of a truncated power law summed with two Gaussian peaks and multiplied by a low-edge smoothing factor.
We include this model both to compare it with the semiparametric one and to validate our code against LVK cosmological pipelines \citep{Mastrogiovanni:2023emh, Gray:2023wgj}.
The semiparametric model is built as \citep{Edelman:2021zkw}
\begin{equation}
    p(m_1 \mid \boldsymbol{\lambda}_m) \propto \mathcal{SP}(m_1; \alpha, m_{\rm low}, m_{\rm high}, \delta_m) \exp\l( s^{(d)}(m_1; \boldsymbol{c}, \boldsymbol{k})\r),
\end{equation}
where $ \mathcal{SP}(m_1; \alpha, m_{\rm low}, m_{\rm high}, \delta_m)$ is a power law truncated in the range $[m_{\rm low}, m_{\rm high}]$, smoothed at the lower edge, and with spectral index $-\alpha$ \citep[see Eqs. C25-C26 of][]{LIGOScientific:2025jau}, and $s^{(d)}(m_1)$ is a \bspline defined as
\begin{equation}
    s^{(d)}(m_1; \boldsymbol{c}, \boldsymbol{k}) = \sum_{i=1}^{N_{\rm coeff}} c_i \cdot B_{d,i}(m_1; \boldsymbol{k})
\end{equation}
Here, $B_{d,i}(m_1; \boldsymbol{k})$ are the spline basis functions of degree $d$ defined recursively from the knot sequence $\boldsymbol{k} = \{k_1, k_2, \dots, k_{N_{\rm knots}}\}$ using the Cox-de Boor formula, and $\{c_i\}$ are the spline coefficients that scale each basis function. 
These are the free parameters of the \bspline that flexibly control deviations from the underlying truncated power law.
In the semiparametric model, the secondary mass distribution, $p(m_2 \mid m_1, \boldsymbol{\lambda}_m)$, is modeled as in the \pltwop one: a smoothed power law with slope $\beta$, truncated in the interval $[m_{\rm low}, m_{\rm high}]$, with $m_2$ constrained to be $m_2\leq m_1$.

\begin{figure}[!htb]
    \centering
    \resizebox{0.88\hsize}{!}{\includegraphics{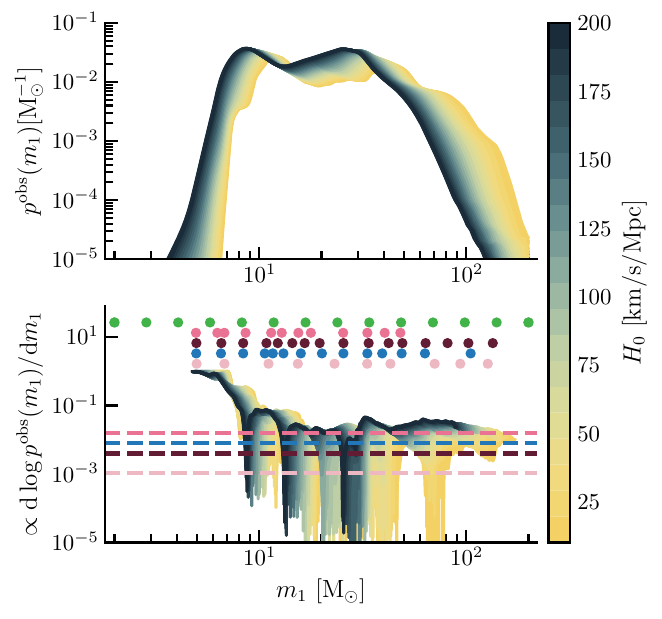}}
    \caption{Mean observed source-frame primary MD for different $H_0$ values (top) and its log-derivative (bottom) used to determine knot positions (dots) at different thresholds (dashed lines). Green points are knots for the \textsc{pls-log-14} model.}
    \label{fig:spline-knots-algo}
\end{figure}

Knot positions are fundamental as they determine the total number of spline coefficients, $N_{\rm coeff} = N_{\rm knots} + d - 1$, and define the Greville abscissae as the averages of consecutive $d+1$ knots. 
The latter corresponds to the nodes in the $m_1$-space where $p(m_1)$ can deviate from the baseline power law, according to the values of $\{c_i\}$.
In this work, we consider cubic, $d = 3$, \bsplines, and we explored different knot configurations.
In one configuration, knots are logarithmically spaced \citep[as in][]{Edelman:2021zkw} across the whole $m_1$-interval that can be explored within the prior range considered (see \Cref{tab:priors}).
In the other configurations, spline-knot positions are determined with a novel data-driven procedure that captures the full complexity of the MD while avoiding an unnecessary increase in knot number.
For a given value of $H_0$, we computed the mean observed source-frame primary mass distribution of all GW events (top panel of \Cref{fig:spline-knots-algo}). 
We then identified the knot positions for each specific $H_0$ value as the points of highest variation in this distribution, corresponding to the peaks of the derivative of its logarithm (bottom panel of \Cref{fig:spline-knots-algo}).
This procedure is repeated for various $H_0$ values drawn from its prior, resulting in a collection of possible knots. 
The final set of knots is identified using a clustering algorithm, which fits the collection of knots with a series of Gaussian mixture models characterized by a different number of components. 
The optimal number of clusters is found by minimizing the Bayesian information criterion, which effectively penalizes model complexity to prevent overfitting. 
The final knot positions are the centers of the resulting clusters.
In \Cref{fig:ppc-2}, we check that this data-driven procedure does not impose mass distribution features that could affect cosmological constraints a priori.

The number of knots is determined by the threshold used in the peak-finding algorithm.
We examined four different thresholds, shown in \Cref{fig:spline-knots-algo}, which produce four sets of knots: 10 (\textsc{pls-dd-10}), 12 (\textsc{pls-dd-12}), 14 (\textsc{pls-dd-14}), and 16 (\textsc{pls-dd-16}).
For the logarithmically spaced configuration, we use 14 knots (\textsc{pls-log-14}) for a direct comparison with the data-driven \textsc{pls-dd-14} case.
A Gaussian prior $\mathcal{G}_{\mu,\sigma}$ with mean $\mu = 0$ is imposed on each spline coefficient.
For \textsc{pls-dd-14}, we test several values of standard deviation $\sigma$: $0.5$, $1$, $2$, $3$, and $5$.
In all other cases, $\sigma$ is fixed to $2$, which yielded the best results in \textsc{pls-dd-14} (see \Cref{sec:results}).
To sample the likelihood implemented in \texttt{CHIMERA}, we used \texttt{pocoMC} \citep{Karamanis:2022alw, Karamanis:2022ksp}, an adaptive sequential Monte Carlo sampler that estimates the posterior distribution and the evidence of each model.

\section{Results}\label{sec:results}

\subsection{Model comparison}
We quantitatively compared the results obtained with different models using the Bayes factor (BF), computed relative to the \textsc{pl2p} baseline, and the Deviance information criterion (DIC). 
A BF > 20 (150) indicates strong (very strong) evidence for a model over \textsc{pl2p}, while a $\Delta\text{DIC}$ > 6 (relative to the model with the lowest DIC) suggests a substantially worse fit  \citep{Kass01061995,Rezaei:2021qpq}.
The results are summarized in \Cref{tab:summary_results}, where we also present posterior predictive checks for each model.
According to the BF criterion, none of the PLS models are disfavored relative to \textsc{pl2p}. 
A narrow Gaussian prior ($\sigma = 0.5$) or fewer knots ($N = 10, 12$) penalize the emergence of structures in the MD, and are basically equivalent to the \textsc{pl2p} (BF $\lesssim 5.3$). 
Stronger evidence emerges when spline coefficients are allowed to vary sufficiently ($\sigma \geq 2$), with the best model being \textsc{pls-dd-14-$\mathcal{G}_2$} (BF = 226). 
A model with the same prior and number of knots, but logarithmically spaced, performs considerably worse (BF = 7.28). 
Adding more knots (\textsc{pls-dd-16-$\mathcal{G}_2$}) does not improve the evidence.
The $\Delta$DIC criterion identifies the most flexible model (\textsc{pls-dd-14-$\mathcal{G}_5$}) as the best one, although all models with $\sigma \geq 2$ are similarly favored over \textsc{pl2p}.
In accordance with the previous criterion, models with few knots, low $\sigma$, or logarithmic knot spacing, show large $\Delta\text{DIC}$ values and are practically indistinguishable from \textsc{pl2p}.

\subsection{Mass distribution constraints}

\begin{figure*}[!htb]
    \centering
    \includegraphics[width=0.95\textwidth]{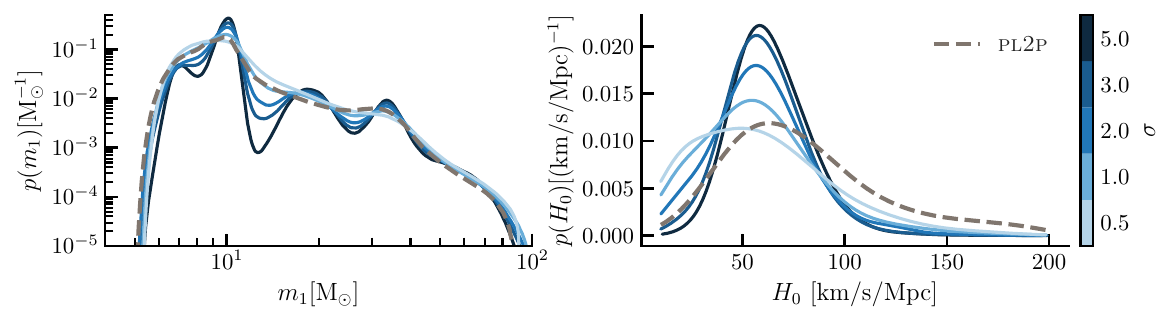}
    \vskip-8pt
    \includegraphics[width=0.95\textwidth]{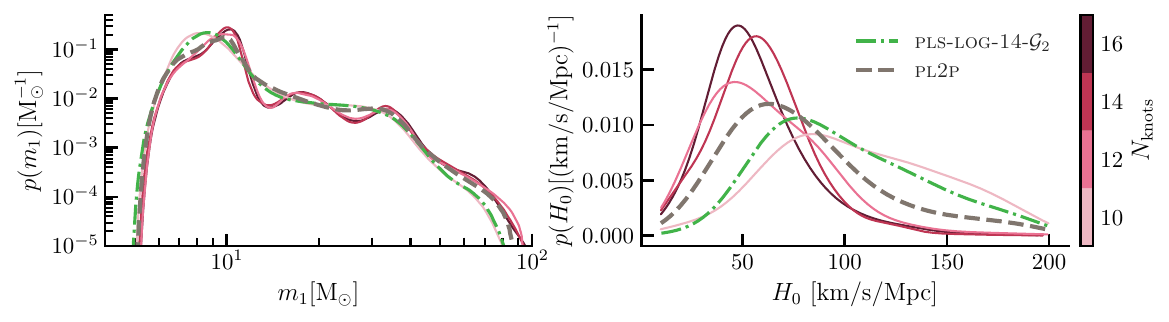}
    \caption{
    Median of the PPD of the primary mass and $H_0$ posterior derived from GWTC-4.0 using different models. Top: Results for the 14 data-driven knots (\textsc{pls-dd-14}) with varying prior widths ($\sigma$) on the spline coefficients. Bottom: the data-driven knot sets with different knot counts $N$ (red curves), alongside the logarithmic case (\textsc{pls-log-14-$\mathcal{G}_2$}; green curve), all using $\sigma = 2$.}
    \label{fig:MF_H0_varying_sigmas_Nknots}
\end{figure*}

We plot the median of the predictive posterior distribution (PPD) for the primary mass for different models in the left panels of \Cref{fig:MF_H0_varying_sigmas_Nknots}, with the corresponding error bars provided in \Cref{app:more_on_mf}.
For the \textsc{pls-dd-14} model, when $\sigma \geq 2$, three distinct peaks emerge at approximately $10$, $18$, and $33\,\mathrm{M}_\odot$. 
These substructures, also reported by \citet{LIGOScientific:2025pvj} and \citet{Tiwari:2025oah} (see \Cref{app:more_on_mf} for a direct comparison), are not captured for $\sigma < 2$.
We also note that the gap between the first and second peaks becomes more pronounced with increasing $\sigma$. 
In contrast, the \textsc{pl2p} model recovers only two broader peaks: around $9.9 \, \mathrm{M}_\odot$ and $31.5 \mathrm{M}_\odot$; the latter seems to smooth out the 18 and 33 $\mathrm{M}_\odot$ peaks observed in the spline model.
Increasing the number of knots reveals a small bump at $\sim 60 \mathrm{M}_\odot$, consistent with findings by \cite{Pierra:2026ffj}, though its presence in our model is not fully evident. 
Conversely, using fewer knots or logarithmically spaced knots produces a fit essentially equivalent to the \textsc{pl2p} model, capturing only the two broader peaks as these models are not flexible enough to capture the structures of $p(m_1)$ suggested by the data.

\begin{figure*}[!htb]
    \centering
    \includegraphics[width=0.95\textwidth]{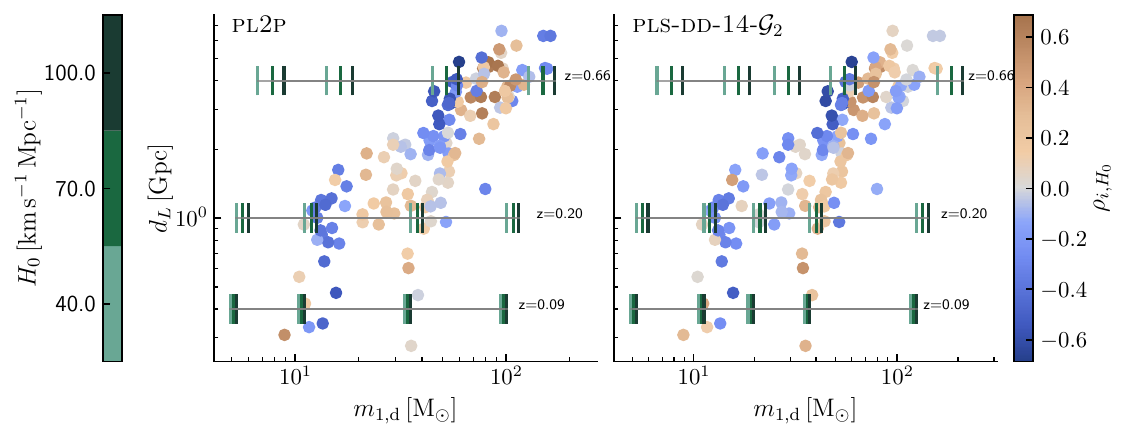}
    \caption{Events of GWTC-4 plotted in the $m_{1,\dd}-d_L$ plane, colored according to the Spearman coefficient between $H_0$ and their single-event marginalized likelihood with two different mass models. The ticks represent the MD features of the different models (minimum and maximum mass, as well as the position of the main peaks), plotted at different $d_L$ and $H_0$ values, as indicated by the green color scale. The luminosity distance values correspond to redshifts $0.09,\,0.2,$ and $0.66$ in a flat $\Lambda$CDM cosmology with $H_0 = 70\si{\kilo\meter\per\second\mega\parsec}$ and $\Omega_{\rm m,0} = 0.3$.}
    \label{fig:constraining-power-pls}
\end{figure*}

\subsection{\texorpdfstring{$H_0$}{H0} constraints}

We find that models that capture substructures in $p(m_1)$ generally lead to tighter constraints on $H_0$, as shown in the right panels of \Cref{fig:MF_H0_varying_sigmas_Nknots}. 
The model \textsc{pls-dd-14-$\mathcal{G}_2$}, favored by the BF, yields 
\begin{equation}
    H_0 = 57.8^{+21.9}_{-20.6} \, \mathrm{km/s/Mpc},
\end{equation}
an improvement of $\sim12\%$ over \textsc{pl2p}. 
The tightest constraint comes from \textsc{pls-dd-14-$\mathcal{G}_5$}, which is the model with the lowest DIC and that exhibits more pronounced features in $p(m_1)$.
This model gives 
\begin{equation}
    H_0 = 61.7^{+19.3}_{-14.9} \, \mathrm{km/s/Mpc},
\end{equation}
corresponding to a $\sim21\%$ improvement relative to \textsc{pl2p}.
Conversely, when fewer knots are used or when they are spaced logarithmically across the full prior range, the constraints on $H_0$ are much weaker, as not all the MD features are captured.

\subsection{Constraining power and correlations}

To determine which events are most informative for $H_0$, we computed the Spearman correlation coefficient $\rho$ between the $H_0$ samples and the corresponding marginal likelihood of each event.
This coefficient quantifies how strongly the single-event likelihood correlates with $H_0$, offering a measure of each event's individual constraining power on $H_0$.
Results obtained with the \pltwop and \textsc{pls-dd-14}$ (\sigma=2)$ models are shown in \Cref{fig:constraining-power-pls}.
The plot displays the events in the detector-frame primary mass - luminosity distance ($m_{1,\dd}-d_L$) plane, with points colored according to their Spearman coefficient.
We also show how the main MD features captured by the two models evolve with $H_0$: the power law edges, $m_{\rm low},\,m_{\rm high}$, and the peaks (two for \pltwop and three for \textsc{pls-dd-14-}$\mathcal{G}_2$).

In the \textsc{pls-dd-14-}$\mathcal{G}_2$ case, events with detector-frame primary mass in the range $[20, 35]\,\msun$ anticorrelate with $H_0$.
In contrast, the same events in the \pltwop show a correlation with the Hubble constant. 
These events correspond to those near the additional peak in the source-frame MD at $18\,\msun$.
In both fits, a clear anticorrelation-correlation pattern is present between $H_0$ and events located before and after the main peaks of the MD at $10$ and $31-33\,\msun$ (in source-frame). 
Interestingly, the \textsc{pls-dd-14-}$\mathcal{G}_2$ fit also shows an additional strong pattern across the position of the third identified peak, demonstrating that it carries additional cosmological information.

\begin{figure}[!htb]
    \centering
    \resizebox{0.9\hsize}{!}{\includegraphics{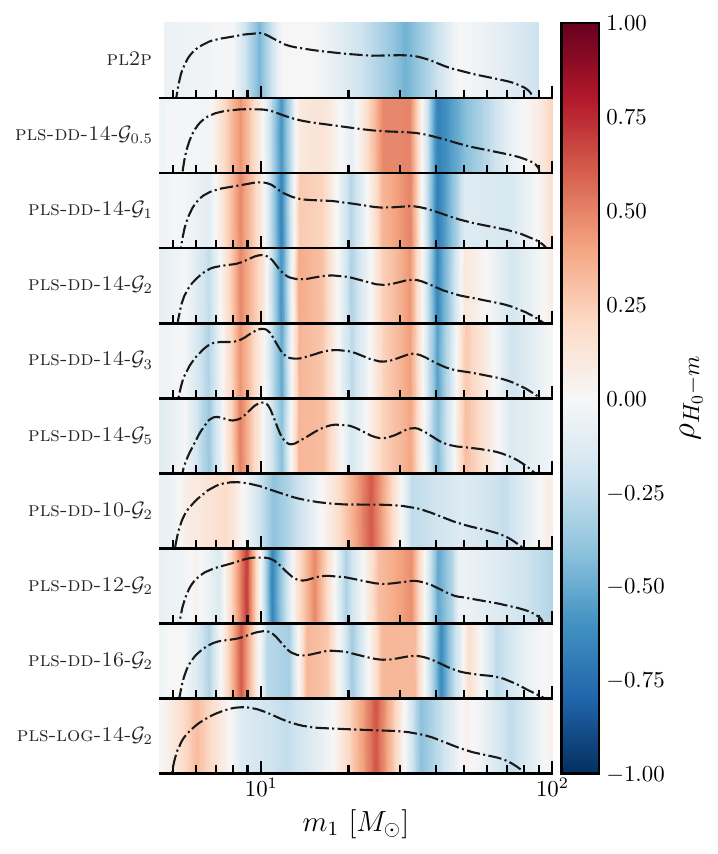}}
    \caption{Spearman correlation coefficients between $H_0$ and the parameters of $p(m_1)$ (medians in black).}
    \label{fig:corrrelations}
\end{figure}
To study how the parameters of the MD correlate with $H_0$, we computed Spearman correlation coefficients between $H_0$ and the parameters describing $p(m_1)$, that are $m_{\rm low}$ and $m_{\rm high}$ (all models), Gaussian peak positions (\textsc{pl2p}), and the spline coefficients (\textsc{pls} models). 
The results are shown in \Cref{fig:corrrelations}.
In \textsc{pl2p}, the two peaks are slightly anticorrelated with $H_0$.
In \textsc{pls-dd} models with $N\geq12$, spline coefficients before each peak correlate with $H_0$, while those after anticorrelate.
For \textsc{pls-dd-10} and \textsc{pls-log-14}, this pattern is less clear due to insufficient nodes near the peaks, thus proving the importance of placing knots around the relevant structure of the mass distribution.
The strongest correlations with $H_0$ come from spline coefficients near the first peak. 
However, in best-performing models, \textsc{pls-dd-14} with $\sigma = 2$ and $5$, the second and third peak also correlate, almost equally between each other, with $H_0$.

\section{Conclusions}
In this work, we derived constraints on cosmological and population parameters from a subset of GWTC-4.0 BBHs.
We adopted a semiparametric mass model based on \bsplines.
We showed that when the spline coefficients are given sufficient freedom, the model captures more substructures in the mass distribution than simpler parametric models.
Using statistical tests, we confirmed that capturing such substructures gives tighter constraints on $H_0$, yielding improvements of $12\%$ to $21\%$ for the models favored by BF and DIC.
We also presented a data-driven method to place spline knots efficiently around features in the observed mass distribution. 
This avoids the computational cost of adding dozens of knots and keeps the model dimensionality manageable.
This method still depends on a few hand-tuned parameters, whose effect has been investigated in this work. 
A future extension of this work could include more optimized data-driven methods for defining spline knot positions, or the use of a larger number of knots combined with a smoothness prior to prevent overfitting \citep{Edelman:2022ydv}.
Another natural next step is to extend the \bspline method to the mass distribution of all compact binaries coalescences, not only of BBH.
Overall, our work demonstrates the importance of accurately modeling all substructures in the mass distribution to increase the constraining power on cosmological parameters, a crucial step to exploit spectral sirens as a robust and precise cosmological probe.

\begin{acknowledgements}
    We acknowledge the ICSC for awarding this project access to the EuroHPC supercomputer LEONARDO, hosted by CINECA (Italy).
    This material is based upon work supported by NSF's LIGO Laboratory which is a major facility fully funded by the National Science Foundation.
    MT acknowledges the funding from the European Union - NextGenerationEU, in the framework of the HPC project – “National Center for HPC, Big Data and Quantum Computing” (PNRR - M4C2 - I1.4 - CN00000013 – CUP J33C22001170001). MM acknowledges the financial contribution from the grant PRIN-MUR 2022 2022NY2ZRS 001 “Optimizing the extraction of cosmological information from Large Scale Structure analysis in view of the next large spectroscopic surveys” supported by NextGenerationEU. MM and NB acknowledge the financial contribution from the grant ASI n. 2024-10-HH.0 “Attività scientifiche per la missione Euclid – fase E”.
\end{acknowledgements}

\bibliographystyle{aa} 
\bibliography{ref.bib}

\begin{appendix}
\nolinenumbers

\section{Priors and statistical significance of the results}\label{app:statistical-summary}
\begin{figure}[!htb]
    \centering
    \includegraphics[width=\linewidth]{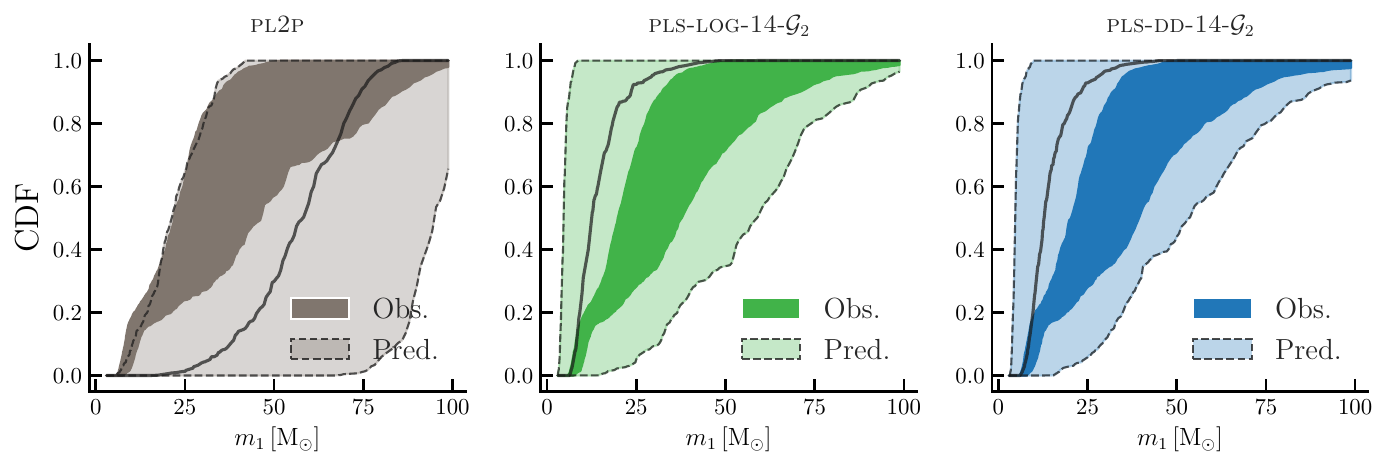}
    \caption{Prior predictive observed primary-mass cumulative distributions, compared with observed ones.}
    \label{fig:ppc-2}
\end{figure}

In \Cref{fig:ppc-2} we show the prior predictive checks for some models studied in this work.
This test is crucial to assess if the data-driven procedure used to determine spline-knot positions artificially adds cosmological information.
In this test, hyperparameters are drawn from the prior and propagated through the full forward model, including selection effects, to generate the cumulative distributions of the predicted observed primary-mass.
These are then compared with the empirically observed ones, in which we also have to correct the GW selection effects using injections.
We find that, for both spline models, the prior predictive contours remain broad and do not show any mass features a priori. 
In particular, the model with data-driven knots shows a pattern completely comparable with the one obtained from the log-spaced model, where the knots' placement is not informed by the data.
This indicates that the knot placement primarily provides adaptive resolution rather than artificially adding cosmological information.

\begin{figure*}[b!]
    \centering
    \includegraphics[width=\textwidth]{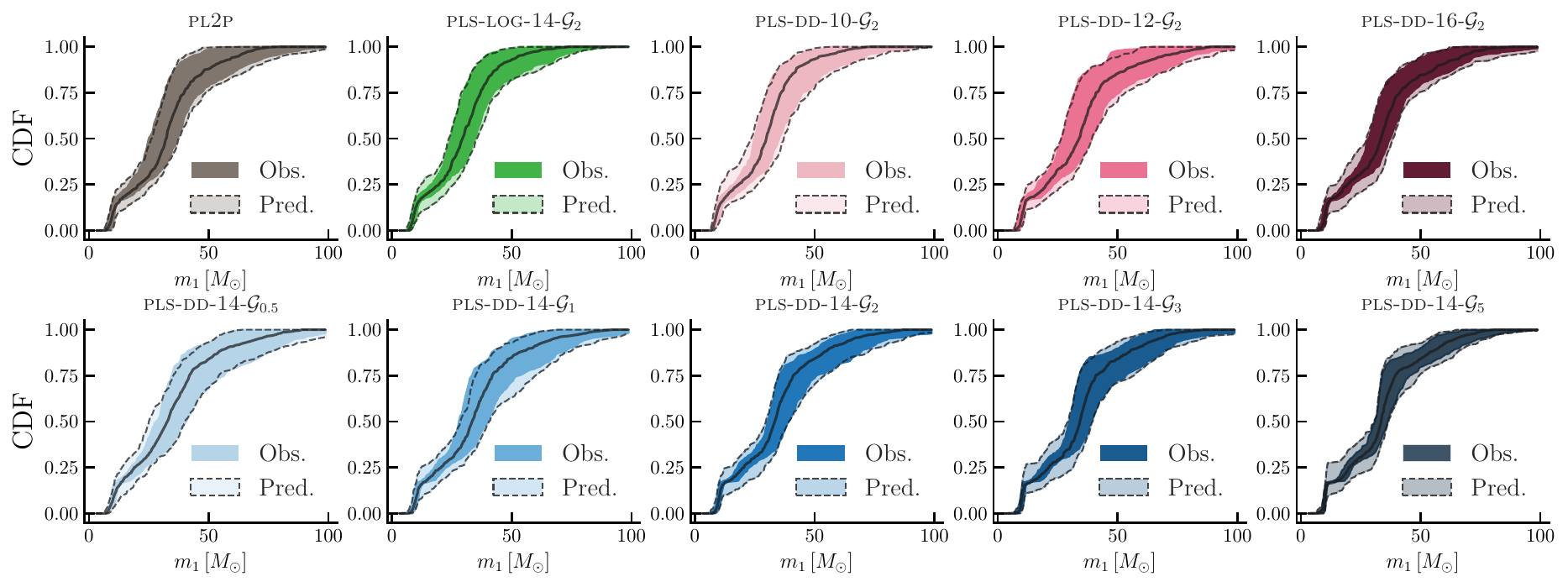}
    \caption{Posterior predictive check for all the models tested in this work. In darker colors are shown the observed cumulative distribution of the BBH population, while in lighter colors the predicted one, given the considered models (reported at the top of each panel). Solid black lines are the medians of the predicted CDFs of $m_1$.}
    \label{fig:PPD}
\end{figure*}

In \Cref{tab:priors} we summarize the population and cosmological parameters, and the relative priors adopted in this work. 
Here, we used a new version of the \chimera pipeline, which we release along with this paper.\footnote{The updated \chimera version is publicly available at \url{https://github.com/cosmoStatGW/chimera}}
In \Cref{tab:summary_results} we present the constraints on $H_0$, the BF, and the $\Delta DIC$ for each model studied in this work.
In \Cref{fig:PPD} we show the predictive posterior check (PPC) for each model that we explored.
In particular, we plot both the predicted and observed cumulative distribution functions (CDF) of $m_1$ for each model considered.
Overall, all models produce an observed distribution that is compatible with the predicted one, thus proving that none of them is ruled out based only on the PPC. However, as discussed in the letter, the BF and $\Delta$DIC clearly show that simpler models are disfavored by data.

\begin{table}[h]
\caption{Summary of the cosmological and population parameters, and relative priors, considered in this work.}
\centering\resizebox{\hsize}{!}{
\begin{tabular}{llcc}
\hline
Symbol & Description & Model & Prior \\
\hline
\multicolumn{4}{l}{\textbf{Cosmology (flat $\Lambda$CDM)}} \\
$H_0$ & Hubble constant [km/s/Mpc] & All & $\mathcal{U}(10.0, 200.0)$ \\
$\Omega_{\rm m,0}$ & Matter energy density & All & Fixed to 0.3065 \\
\hline
\multicolumn{4}{l}{\textbf{Mass distributions}} \\
$\alpha$ & Primary power law slope & All & $\mathcal{U}(1.5, 12)$ \\
$\beta$ & Secondary power law slope & All & $\mathcal{U}(-4, 12)$ \\
$\delta_m$ & Smoothing parameter $[\msun]$ & All & $\mathcal{U}(0.001, 10)$ \\
$m_{\rm low}$ & Power law lower limit $[\msun]$ & All & $\mathcal{U}(2, 10)$ \\
$m_{\rm high}$ & Power law upper limit $[\msun]$ & All & $\mathcal{U}(50, 200)$ \\
$\mu^{\rm low}_g$ & Position of the first Gaussian peak $[\msun]$ & \textsc{pl2p} & $\mathcal{U}(5, 100)$ \\
$\sigma^{\rm low}_g$ & Width of the first Gaussian peak $[\msun]$ & \textsc{pl2p} & $\mathcal{U}(0.4, 5)$ \\
$\mu^{\rm high}_g$ & Position of the second Gaussian peak $[\msun]$ & \textsc{pl2p} & $\mathcal{U}(5, 100)$ \\
$\sigma^{\rm high}_g$ & Width of the second Gaussian peak $[\msun]$ & \textsc{pl2p} & $\mathcal{U}(0.4, 10)$ \\
$\lambda^{\rm low}_g$ & Mixing fraction of the first Gaussian peak & \textsc{pl2p} & $\mathcal{U}(0, 1)$ \\
$\lambda^{\rm high}_g$ & Mixing fraction of the second Gaussian peak & \textsc{pl2p} & $\mathcal{U}(0, 1)$ \\
$c_i$ & Spline coefficients & \textsc{pls} & $\mathcal{G}(\mu = 0, \sigma =0.5, 1, 2, 3, 5)$ \\
\hline
\multicolumn{4}{l}{\textbf{Rate evolution (Madau-like)}} \\
$\gamma$ & Slope at $z<z_p$ & All & $\mathcal{U}(0, 12)$ \\
$\kappa$ & Slope at $z>z_p$ & All & $\mathcal{U}(0, 6)$ \\
$z_{\rm p}$ & Peak redshift & All & $\mathcal{U}(0, 4)$ \\
\hline
\end{tabular}}
\tablefoot{The symbol $\mathcal{U}(\cdot)$ denotes a uniform prior distribution, while $\mathcal{G}$ denotes a Gaussian distribution with mean $\mu$ and standard deviation $\sigma$. }
\label{tab:priors}
\end{table}

\begin{table}[t]
\caption{Median and $68\%$ credible interval of $H_0$, BF, and $\Delta$DIC per model.}
\centering\resizebox{\hsize}{!}{
\begin{tabular}{lccc}
\hline
Model & $H_0\,\mathrm{[km/s/Mpc]}$ & Bayes factor & $\Delta$DIC\\
\hline
\textsc{pl2p} & $72.6^{+42.7}_{-27.5}$ & $1.0$ & $20.1$ \\[2pt]
\textsc{pls-dd-14}-$\mathcal{G}_{0.5}$ & $54.6^{+36.5}_{-30.7}$ & $2.59$ & $18.2$ \\[2pt]
\textsc{pls-dd-14}-$\mathcal{G}_{1}$ & $55.4^{+28.0}_{-26.6}$ & $81.8$ & $9.42$ \\[2pt]
\textsc{pls-dd-14}-$\mathcal{G}_{2}$ & $57.8^{+21.9}_{-20.6}$ & $226$ & $2.65$ \\[2pt]
\textsc{pls-dd-14}-$\mathcal{G}_{3}$ & $59.4^{+19.2}_{-16.2}$ & $134$ & $0.26$ \\[2pt]
\textsc{pls-dd-14}-$\mathcal{G}_{5}$ & $61.7^{+19.3}_{-14.9}$ & $24.1$ & $0.00$ \\[2pt]
\textsc{pls-dd-10}-$\mathcal{G}_{2}$ & $102.1^{+51.2}_{-36.3}$ & $5.27$ & $13.4$ \\[2pt]
\textsc{pls-dd-12}-$\mathcal{G}_{2}$ & $56.8^{+32.5}_{-23.4}$ & $60.8$ & $7.93$ \\[2pt]
\textsc{pls-dd-16}-$\mathcal{G}_{2}$ & $52.0^{+26.2}_{-16.5}$ & $114$ & $5.41$ \\[2pt]
\textsc{pls-log-14}-$\mathcal{G}_{2}$ & $93.0^{+48.6}_{-30.8}$ & $7.28$ & $13.3$ \\[2pt]
\hline
\end{tabular}
}
\label{tab:summary_results}
\end{table}

\section{Mass distribution results}\label{app:more_on_mf}

\begin{figure*}[ht]
    \centering
    \includegraphics[width=\textwidth]{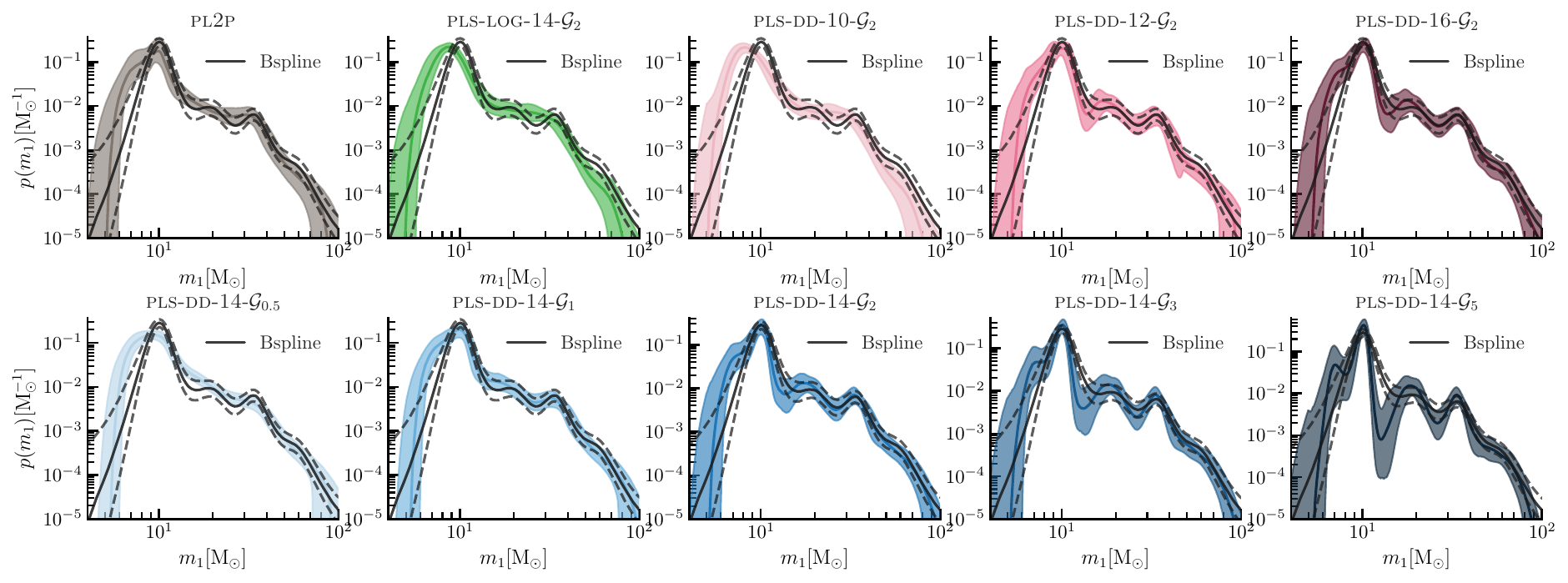}
    \caption{Median and 68\% credible interval of the PPD for the primary mass for each model. We compare these results with those found using another weakly-parametrized approach by \cite{LIGOScientific:2025pvj} (black dashed line).}
    \label{fig:full-mf}
\end{figure*}

In \Cref{fig:full-mf} we show the constraints on the PPD for the primary mass for each model considered in this work. 
In particular, we plot the median and 68\% credible interval of $p(m_1)$ for each model.
In the same figure, we also compare the model studied against results obtained with a different \bspline model by \cite{LIGOScientific:2025pvj}.
In this comparison, it is important to underline that \cite{LIGOScientific:2025pvj} does not vary the cosmological parameters, but fixes them to some fiducial values, and uses a slightly different dataset with 16 more BBHs.
Nevertheless, the comparison is extremely interesting, since we note that the peak at around 20 $\mathrm{M}_\odot$ present in the \bspline model is also derived by the preferred \texttt{pls} models in our analysis. 
We conclude that using \texttt{pls} models with a sufficient number of knots and enough freedom in the spline coefficients leads to a more accurate reconstruction of the mass distribution. 
This, in turn, improves the inferred cosmological parameters, as the additional features enhance the constraining power of the spectral sirens approach.

\end{appendix}
\end{document}